\documentclass[10pt]{article}

\usepackage{epsfig}

\def\lsim{\mathrel{\rlap{\lower4pt\hbox{\hskip1pt$\sim$}}
    \raise1pt\hbox{$<$}}}         %less than or approx. symbol
\def\gsim{\mathrel{\rlap{\lower4pt\hbox{\hskip1pt$\sim$}}
    \raise1pt\hbox{$>$}}}         %greater than or approx. symbol

\def\overleftrightarrow#1{\vbox{\ialign{##\crcr
    $\leftrightarrow$\crcr
    \noalign{\kern 1pt\nointerlineskip}
    $\hfil\displaystyle{#1}\hfil$\crcr}}}

\newcommand{\pp}{\\[0.5cm]}
\newcommand{\be}{\begin{equation}}
\newcommand{\ee}{\end{equation}}
\newcommand{\bea}{\begin{eqnarray}}
\newcommand{\eea}{\end{eqnarray}}

\newcommand{\vk}{k}

\begin{document}

\hfill {\bf TUM/T39-02-19} \\
\vspace{0.01in}
\hfill {\bf ECT*-02-26} \\
\vspace{0.1in}

\begin{center}
{\bf \Large A Note on Asymptotic \\ Freedom at High Temperatures\footnote{Work
supported in part by BMBF, GSI and by the European Commission under contract HPMT-CT-2001-00370.}}
\end{center}

\begin{center}
{R.A. Schneider}\\
{\small \em  Physik-Department, Technische Universit\"{a}t M\"{u}nchen, 85747 Garching,
Germany\\}
{\small \em  ECT*, Villa Tambosi, 38050 Villazzano (Trento), Italy\\}
%
%\maketitle
\end{center}

\begin{abstract}
This short note considers, within the external field approach outlined in \cite{Schneider:2002}, the role of the lowest lying gluon Landau mode in QCD in the high temperature limit. Its influence on a temperature- and field-dependent running coupling constant is examined. The thermal imaginary part of the mode is temperature-independent in our approach and exactly cancels the well-known zero temperature imaginary part, thus rendering the Savvidy vacuum stable. Combining the real part of the mode with the contributions from the higher lying Landau modes and the vacuum contribution, a field-independent coupling $\alpha_s(T)$ is obtained. It can be interpreted as the ordinary zero temperature running coupling constant with average thermal momenta  $\langle k \rangle \approx 2 \pi T$ for gluons and $\langle k \rangle \approx \pi T$ for quarks.   
\end{abstract}
\vspace{0.25in}
In \cite{Schneider:2002}, we have calculated an effective charge $\alpha^{\rm eff}_s$ in QCD, extending the approach of refs.\cite{NKN, JLP} to finite temperature. Instead of a loop expansion of the gluon self-energy, the thermal energy shift
\be
\Delta E(T, H) = - \frac{1}{2} \left[ 4\pi \chi(H, T) \right] VH^2 - E(T)_{\rm vac}, \label{master1}
\ee
 of the perturbed thermal vacuum (the ``Savvidy'' vacuum) to order $\alpha_s$, after applying an external
static chromomagnetic field $H$, was evaluated at temperatures $T \gg \Lambda_{\rm QCD}$. Identifying $2eH$ with the scale $k^2$ at which the system is probed, as at $T=0$ ($e$ is the product of the strong coupling constant $g$ times the charge number $q$ that involves the structure constants of SU($N_c$)), we have investigated the high temperature limit
\be
\frac{eH}{T^2} \equiv \delta^2 \ll 1. \label{high_T}
\ee
Note that condition (\ref{high_T}) is equivalent to the hard thermal loop (HTL) approximation in thermal perturbation theory at the one-loop level, where the external momentum $k$ that flows into a loop is taken to be 'soft' as compared to the 'hard' internal thermal scale $T$, i.e. $k^0, |\vec{k}| \ll T$. Subsequently, we have extracted a temperature- and momentum-dependent dielectric permittivity $\epsilon(\vk, T)$ by use of the relation
\be
\alpha^{\rm eff}_s(\vk, T) \equiv \frac{\alpha_s}{\epsilon(\vk, T)} = \frac{\alpha_s}{1 - 4\pi \chi(\vk,
T)}.\label{eps_mu_T}
\ee
Here, $\chi(\vk, T) = \chi_g + \chi_q$ is the sum of gluon and quark magnetic susceptibilities. An expression similar to (\ref{master1}) exists at zero temperature and has to be added to (\ref{master1}) in order to obtain the total result, so  $\chi_{\rm total} = \chi_0(k, \Lambda) + \chi(k, T)$, where the  zero temperature susceptibility $\chi_0(k, \Lambda)$ has been calculated in \cite{NKN, JLP}. $\Lambda$ corresponds to a zero temperature scale that characterises the ``medium'' and is ultimately identified with the renormalisation reference point of the coupling.
\pp
For the gluonic energy difference $\Delta E_g(T)$, the result reads for a gluon of charge $q$ \cite{Schneider:2002} 
\be
\Delta E_g(T, H)  =  VT^4 \left\{ - \frac{\delta^2}{6} - \frac{5 \delta^4}{48\pi^2} \left[ \log \left( \frac{\delta^2}{16\pi^2} \right) + 2\gamma + \frac{11}{5} \log 3 - \frac{13}{10}  \right] \right\}. \label{Delta_E}
\ee
With eqs.(\ref{master1}) and (\ref{eps_mu_T}), the $- \delta^2/6$ term leads to a Landau pole in the infrared, as also found in more sophisticated renormalisation group analyses of the running coupling at finite temperature. This feature is unphysical since it would signal antiscreening of colour over large distances, which is in contrast to expectations from asymptotic freedom and lattice calculations \cite{FK}. In this short note, we show that the Landau pole vanishes with the inclusion of the lowest lying gluonic Landau mode in eq.(\ref{Delta_E}) and, furthermore, that all dependence on $H$ drops out when $\chi(H, T)$ is combined with the zero temperature result. In addition, the imaginary part of that Landau mode exactly cancels the zero temperature imaginary part of $\Delta E$ as found in \cite{NO}, rendering the vacuum stable, in contrast to previous approaches.  
\pp
The calculation of $\Delta E_g(T)$ involves a sum over all Landau levels
\be
E_{n,k_3, s_3} = \sqrt{k_3^2 + 2eH \left( n + 1/2 + s_3 \right)}, \label{E_gluon}
\ee
weighted by the corresponding thermal occupation probabilities $[\exp(E_{n,k_3, s_3}/T) - 1]^{-1}$. Here, $n$ labels the mode, $k_3$ is the 3-component of the momentum and $s_3 = \pm 1$, the $z$-component of the spin of the transverse gluons. As is well-known, the lowest lying Landau level (LLL) with $n=0$ and $s_3 = -1$ acquires an imaginary part for small $k_3$ already at zero temperature \cite{NO}. This tachyonic instability signals that the constant field $H$ will decay to some new, unspecified vacuum state, maybe accompanied by the formation of a chromomagnetic condensate. In previous approaches, this feature persisted even at high temperatures \cite{NI}, despite asymptotic freedom, but could be avoided by the {\em ad hoc} introduction of some thermal electric ($m_e \sim gT$) \cite{EO} or magnetic gluon mass ($m_m \sim g^2T$) \cite{EP}. The influence of the LLL on the high-temperature physics was at best inconclusive. In \cite{Schneider:2002}, we have therefore discarded the contribution of the LLL in obtaining (\ref{Delta_E}). In the following, we will include in eq.(\ref{Delta_E}) the explicit expression for the LLL, which reads
\be
E_{\rm LLL} = V T^4 \left(\frac{\delta^2}{\pi^2} \int \limits_0^\infty dx \ \frac{\sqrt{x^2 - \delta^2}}{\exp(\sqrt{x^2 - \delta^2}) - 1} \right). \label{E_LLL}
\ee
An expansion of the integral in small $\delta$ (though not a power series) for positive  $x^2 + \delta^2$ exists as
$$
\int \limits_0^\infty dx \ \frac{\sqrt{x^2 + \delta^2}}{\exp(\sqrt{x^2 + \delta^2}) - 1} = \frac{\pi^2}{6} + \frac{\delta^2}{4} \left[ \log \left( \frac{\delta}{4\pi} \right) + \gamma + \frac{1}{2} \right] + \mathcal{O}(\delta^6),  
$$
where $\gamma = 0.5772...$, the Euler-Mascheroni constant. As long as $\delta $ is small in (\ref{E_LLL}), we can analytically continue the expansion to imaginary values of $\delta$. To obtain its correct sign, the usual Feynman $\epsilon$-prescription has to be applied, as done at $T=0$ \cite{NO}: $\delta^2 \rightarrow \delta^2 - i \epsilon$, which leads to $\sqrt{- \delta^2} \rightarrow - i \sqrt{\delta^2}$. 
The only imaginary part at finite temperature then arises from the complex logarithm:
$$
E_{\rm LLL} \simeq  V T^4 \left\{ \frac{\delta^2}{6} - \frac{\delta^4}{8\pi^2} \left[ \log \left( \frac{\delta^2}{16\pi^2} \right)  + 2 \gamma + 1 \right] + i \left( \frac{\delta^4}{8 \pi} \right) \right\}.
$$
Obviously, the positive sign of the imaginary part would indicate a blow-up of the configuration, not a decay, which is unphysical. However, when re-writing $\delta$, the imaginary part 
\be
\mbox{Im}E_{\rm LLL} = \mbox{Im} [ \Delta E(T) ] = +  V \frac{(eH)^2}{8\pi} \label{ImE}
\ee
becomes independent of temperature. At zero temperature, the imaginary part of the energy difference is calculated to be Im$ [ \Delta E_0 ] = - V (eH)^2 / (8 \pi)$ \cite{NO}, which is exactly the opposite of eq.(\ref{ImE}). Taking both contributions into account, the total imaginary part of the energy difference hence vanishes, which renders the Savvidy high temperature vacuum stable. This result has been long sought after, but previous approaches \cite{NI, PER} always found a remaining imaginary part of the form Im$[ \Delta E(T) ]  = - V T^4 [ \delta^3 / (2\pi) ]$. All contributions of order $\delta^3$ in $\Delta E(T)$ cancel, however, within our approach when all contributions up to order $g^2$ are consistently taken into account. In addition, whereas terms proportional to the squared charge $e^2$ do not depend on the direction of the external field $H$ in colour space, terms not quadratic in the coupling, like $\delta^3 \sim e^{3/2}$, are not group-invariant and do depend on the specific colour choice of the magnetic field \cite{PER}, a result that is probably unphysical.
\pp
Combining now the real part of $E_{\rm LLL}$ with the sum over all higher lying Landau modes, eq.(\ref{Delta_E}), we find that  the troublesome $- \delta^2/6$ term is exactly cancelled by Re$E_{\rm LLL}$ which henceforth eliminates the Landau pole in the infrared. Furthermore, the logarithms combine such as to yield, after summing over the colour charges $q^2$ of the adjoint representation,
$$
\Delta E_g = - \frac{1}{2} V H^2 \left[ g^2 \frac{11 N_c}{48 \pi^2} \log \left( \mathcal{A}_g \frac{2eH}{T^2} \right) \right]
$$
with $\mathcal{A}_g = \exp(2\gamma + \log 3 - 1/22)/(32\pi^2)$. The expression in square brackets already stands for $4 \pi \chi_g(H, T)$. Together with the quark contribution $\chi_q(H, T)$ from \cite{Schneider:2002},
$$
4\pi \chi(H, T) = g^2 \frac{11 N_c}{48 \pi^2} \log \left( \mathcal{A}_g \frac{2eH}{T^2} \right) -  g^2 \frac{2 N_f}{48 \pi^2} \log \left( \mathcal{A}_f \frac{2eH}{T^2} \right)
$$
($\mathcal{A}_f = \exp(2\gamma -1) / \pi^2$), which looks deceptively similar to the running coupling at $T=0$.  Indeed, consistently taking into account the zero temperature QCD susceptibility $\chi_0(H, \Lambda)$ \cite{NKN, JLP}, as mentioned above, we arrive at the final expression
\be
\alpha_{s}^{\rm eff}(H, T, \Lambda) =  \alpha_{s}^{\rm eff}(T, \Lambda) = \frac{\alpha_s(\Lambda)}{1 + {\displaystyle \frac{\alpha_s(\Lambda)}{12\pi} \left[ 11 N_c \log \left(\frac{[\bar{\mathcal{A}}_g 2 \pi  T]^2}{\Lambda^2} \right) - 2 N_f  \log \left(\frac{[\bar{\mathcal{A}}_f \pi T]^2}{\Lambda^2} \right) \right] }}, \label{alpha_s_T}
\ee
with $\bar{\mathcal{A}}_g = \exp(- \gamma - 1/2 \log(3/8) + 1/44) \approx 0.938$ and $\bar{\mathcal{A}}_f = \exp(- \gamma + 1/2) \approx 0.926$. Eq.(\ref{alpha_s_T}) looks very similar to what was already found in the QED case in \cite{Schneider:2002}: all field (or momentum) dependence in $\alpha_s^{\rm eff}$ has dropped out. It is just the one-loop running coupling constant at zero temperature, where the loop particles carry some average thermal momentum $\langle k \rangle = \mathcal{O}(T)$, as originally put forward in \cite{CP}. However, we are now in a position to refine that result: coming from the lowest Matsubara frequencies, quarks propagating in a thermal loop should have $\pi T$ as momentum, whereas the lowest non-vanishing bosonic frequency is $2\pi T$ -- and these values are indeed very close to the numbers appearing in the logarithms of (\ref{alpha_s_T}) in front of $T$. Note that this result arises only when one consistently sums up all terms to order $g^2$. Amusingly, the HTL expressions for (\ref{eps_mu_T}) can be recovered from eq.(\ref{master1}) in a cruder approximation omitting  $\mathcal{O}(g)$ contributions to the single Landau levels and the density of states \cite{Schneider:2002}. Extending the calculation to finite quark chemical potential will shed more light on the connection between the presented approach and HTL perturbation theory. 
\pp
We have therefore been able to show for the first time in a self-contained calculation that, for long wavelength modes, the running coupling at finite temperature becomes very simple, as has been expected for long on a phenomenological basis \cite{CP}: it follows from the zero temperature renormalisation group equations, with the momentum scale replaced by a suitable thermal scale that seems to be set by the lowest, non-zero Matsubara frequencies and is then, of course, different for quarks and gluons. This distinction of thermal momentum scales for fermions and bosons is a new outcome of our calculation. The same formal result $\alpha^{\rm eff} = \alpha(\langle k \rangle \simeq \pi T)$ was already found in QED \cite{Schneider:2002}, supporting the setup of the calculation. All collective medium effects over large distances $R \gg 1/T$ at the one-loop level can therefore be subsumed in a running coupling constant that does not depend on $R$. In addition, the thermal imaginary part of the LLL is exactly cancelled by its well-known zero temperature counterpart within our approach, which eliminates the instability of the Savvidy vacuum, a welcome feature.


\begin{thebibliography}{99}

\bibitem{Schneider:2002} R. A. Schneider, Phys. Rev. {\bf D66} (2002) 036003.

\bibitem{NKN} N. K. Nielsen, Am. J. Phys. {\bf 49} (1981) 1171.

\bibitem{JLP} J. L. Petersen, in {\em Proceedings of the 1997 European Summer School of High-Energy Physics}
(Edts. N. Ellis and M. Neubert), CERN 98-03.

\bibitem{FK} F. Karsch, Nucl. Phys. {\bf A698} (2002) 199.

\bibitem{NO} N. K. Nielsen and P. Olesen, Nucl. Phys. {\bf B144} (1978) 376.

\bibitem{NI} M. Ninomiya and N. Sakai, Nucl. Phys. {\bf B190} (1981) 316.

\bibitem{EO} K. Enqvist and P. Olesen, Phys. Lett. {\bf B329} (1994) 195.

\bibitem{EP} P. Elmfors and D. Persson, Nucl. Phys. {\bf B538} (1999) 309. 

\bibitem{PER} D. Persson, Annals Phys. {\bf 252} (1996) 33.

\bibitem{CP} J. C. Collins and M. J. Perry, Phys. Rev. Lett. {\bf 34} (1975) 1353.

\end{thebibliography}
\end{document}